\documentclass{aa}

\usepackage{color}
\usepackage{graphicx,txfonts,natbib,color}
\bibpunct{(}{)}{;}{a}{}{,} 
\title{Siphon flow in a cool magnetic loop} 
\subtitle{}

\author{C. Bethge\inst{\ref{KIS},}\inst{\ref{HAO}} 
\and C. Beck\inst{\ref{IAC},}\inst{\ref{ULL}}
\and H. Peter\inst{\ref{MPS}} 
\and A. Lagg\inst{\ref{MPS}} 
}

\institute{Kiepenheuer-Institut f\"ur Sonnenphysik, Sch\"oneckstr. 6, 79104 Freiburg, Germany\label{KIS}
\and High Altitude Observatory, National Center for Atmospheric
Research\thanks{The National Center for Atmospheric Research is 
  sponsored by the National Science Foundation.}, P.O. Box 3000,
Boulder, CO 80307, USA\newline\email{bethge@ucar.edu}\label{HAO} 
\and Instituto de Astrof{\'\i}sica de Canarias (CSIC), Via Lactea, E-38205 La Laguna, Tenerife, Spain\label{IAC}
\and Departamento de Astrof{\'i}sica, Universidad de La Laguna, E-38206 La Laguna, Tenerife, Spain\label{ULL}
\and Max-Planck-Institut f\"ur Sonnensystemforschung, 37191 Katlenburg-Lindau, Germany\label{MPS}}

\date{Received 26 October 2011 / Accepted 22 November 2011}
 
\abstract{{Siphon flows that are driven by a gas pressure difference
    between two photospheric footpoints of different magnetic field
    strength connected by magnetic field lines are a well-studied
    phenomenon in theory, but observational evidence is scarce.}} 
  {{We investigate the properties of a structure in the solar
      chromosphere in an active region to find out whether the feature
      is consistent with a siphon flow in a magnetic loop filled with
      chromospheric material.}} 
  {{We derived the line-of-sight (LOS) velocity of several
      photospheric spectral lines and two chromospheric spectral
      lines, \mbox{\ion{Ca}{ii} H} 3968.5~\AA$\;$and \mbox{\ion{He}{i}} 10830~\AA, in
      spectropolarimetric observations of NOAA 10978 done with the
      Tenerife Infrared Polarimeter (\mbox{TIP-II}) and the POlarimetric LIttrow
      Spectrograph (POLIS). The structure can be clearly traced in the LOS
      velocity maps and the absorption depth of \mbox{\ion{He}{i}}. The
      magnetic field configuration in the photosphere is inferred
      directly from the observed Stokes parameters and from inversions
      with the H{\tiny E}LI{\tiny X}$^{+}$ code. Data from the full-disk 
      Chromospheric Telescope (ChroTel) in \mbox{\ion{He}{i}}
      in intensity and LOS velocity are used for
      tracking the temporal evolution of the flow, along with TRACE
      \mbox{\ion{Fe}{ix/x}} 171~\AA$\;$data for additional information about
      coronal regions related to the structure under investigation.}} 
 {{The inner end of the structure is located in the penumbra of a
     sunspot. It shows downflows whose strength decreases with decreasing height
     in the atmosphere. The flow velocity in \mbox{\ion{He}{i}} falls
     abruptly from above 40~km~s$^{-1}$ to about zero further into the
     penumbra. A slight increase of emission is seen in the
     \mbox{\ion{Ca}{ii} H} spectra at the endpoint. At the outer end of the
     structure, the photospheric lines that form higher up in the
     atmosphere show upflows that accelerate with height. The
     polarization signal near the outer end shows a polarity opposite
     to that of the sunspot, the magnetic field strength of 580~G is
     roughly half as large as at the inner end. The structure exists
     for about 30 minutes. Its appearance is preceeded by a
     brightening in its middle in the coronal TRACE data.}} 
  {{The observed flows match theoretical predictions of
      chromospheric and coronal siphon flows, with accelerating
      upflowing plasma at one footpoint with low field strength and
      decelerating downflowing plasma at the other end. A tube shock
      at the inner end is probable, but the evidence is not
      conclusive. The TRACE data suggest that the structure forms
      because of a reorganization of field lines after a reconnection
      event.}}  


\begin{document}
\maketitle
\section{Introduction}
Siphon flows in magnetic flux tubes were suggested first by
\cite{Meyer+Schmidt:1968}. Since then they have been addressed
theoretically for the most part, both in the upper solar atmosphere  
\citep[e.g.,][]{Cargill1980,Noci1981} and in the photosphere
\citep[e.g.,][]{Thomas1988,Degenhardt1989,Thomas1991}. These flows are driven
by a difference in gas pressure between the footpoints of the flux
tube caused by a difference of the magnetic field strength. If one
assumes for simplicity a static field-free atmosphere
outside of the flux tube footpoints, the pressure balance can be
described as 
\begin{equation} 
p_e = p_i + \frac{{B_i}^2}{2\mu} \label{eq_siphon}
\end{equation}
at the location of the footpoints, with $p_i$ being the internal gas
pressure, ${B_i}^2/2\mu$ the magnetic pressure inside the flux tube, and
$p_e$ the external gas pressure. An increased magnetic field strength
in one of the footpoints ($B_{\rm inner} > B_{\rm outer}$) therefore
leads to an imbalance in gas pressure, which can then drive a plasma
flow along the flux tube. 

Observational evidence of such siphon flows is, however, scarce:
\cite{Rueedi1992} report a siphon flow signature in
\mbox{\ion{Fe}{i}} 15648~\AA$\;$and \mbox{\ion{Fe}{i}} 15653~\AA$\;$across the
polarity-inversion line of an active region. For the same observation, \cite{Degenhardt1993}
presented evidence for both the siphon flow and a standing tube
shock in the flow based on a comparison of the observed line profiles
with synthetic line profiles from computed siphon flows. \cite{Uitenbroek2006}
identified a siphon flow in \mbox{\ion{Ca}{ii}} 8542~\AA$\;$near a
pore with downflow velocities up to 27~km~s$^{-1}$. A similar siphon
flow in \mbox{\ion{Ca}{ii}} 8542~\AA$\;$without a clear outer
footpoint was reported by \cite{Beck2010}. 

All of these observations
employed scans with a slit spectrograph, i.e., the information is
co-temporal only along the slit and not throughout the whole scanned
region. Because the scans were not done repeatedly, there is no
temporal information about the flow, and it can be hard to distinguish
an actual persistent flow from a plasma motion co-moving with the
slit.

\cite{Doyle2006} detected a short-lived siphon flow in \mbox{\ion{N}{v}}
1238~\AA$\;$in a time sequence of slit spectra. The authors
used additional imaging data in \mbox{\ion{C}{iv}} 1550~\AA$\;$and \mbox{\ion{Fe}{ix/x}}
171~\AA$\;$to estimate both the temperature of the plasma and
the duration of the event; the latter was done by tracking
brightenings moving from one footpoint to the other. At the time the
brightening arrived at the slit placed over the `destination'
footpoint, redshifts of about 20~km~s$^{-1}$ were
seen. Nevertheless, because no information about line-shifts was
available in the \mbox{\ion{C}{iv}} 1550~\AA$\;$and \mbox{\ion{Fe}{ix/x}}
171~\AA$\;$data, one cannot necessarily assume that the brightenings are
indeed indicative of moving plasma. 

Siphon flows are frequently found in loops reaching high up into the
corona. When seen close to the limb in the extreme ultraviolet (EUV),
these sometimes show velocities along the loop exceeding the sound
speed \citep[e.g.,][]{Peres:1997,Brekke+al:1997:loops}. In the past,
these siphon flows were used as one possible explanation to
understand why cool loops can be seen far higher than expected from
the pressure scale height
\citep[e.g.,][]{Peres:1997,Aschwanden+al:2001}. Similar to the loop we
  investigate in this study, these were associated with active
  regions, but in contrast to the (most probably low-lying) loop we
  study here, the EUV loops host plasma well above $10^5$\,K and
  easily reach heights of 50~Mm and more. The flows in the coronal
  EUV loops are (most probably) driven by a difference in the
  \emph{coronal} pressure due to asymmetric heating (the heating in one
  leg is stronger than in the other). This scenario is quite different from the
  siphon flows in low-lying photospheric and chromospheric loops as
  originally suggested by \cite{Meyer+Schmidt:1968}, which are driven
  by a \emph{photospheric} pressure imbalance caused by the difference
  of the magnetic field strength.

In the present study, high-resolution spectropolarimetric data in the
\mbox{\ion{He}{i}} 10830~\AA$\;$multiplet, i.e., in the upper chromosphere,
are complemented with spectroscopic data in \mbox{\ion{Ca}{ii} H} in the lower
chromosphere and imaging data from a full-disk filtergraph instrument
observing the spectral range of \mbox{\ion{He}{i}}. Thanks to a tunable
filter, the latter also allows one to infer the line-of-sight (LOS)
velocity of the plasma motions. The high cadence of the full-disk
observations of 30 secs is essential to track the temporal evolution
of the flow. The magnetic field configuration of the structure in
which the siphon flow takes place is deduced from line parameters and
inversions of the spectropolarimetric data. TRACE \mbox{\ion{Fe}{ix/x}}
171~\AA$\;$coronal data are used to investigate the relation to the
upper atmosphere.  

In the following, the observations and data analysis are described
first (Sect.~\ref{sec:oada}), followed by a description of the
morphology of the structure and the dynamics within at high spatial
resolution in Sect.~\ref{sec:hotf}. The same section also addresses
the temporal evolution of the flow. Section~\ref{sec:msotl} deals with
the magnetic field configuration of the structure. The results are discussed in
Sect.~\ref{sec:discussion}. Based on what is seen in the TRACE data in
the beginning and at the end of the flow, we speculate on the causes
of the initiation and termination of the flow in
Sect.~\ref{sec:speculation} before concluding remarks are being made
in Sect.~\ref{sec:conclusions}.  

\section{Observations and data analysis\label{sec:oada}}
Observations in \mbox{\ion{He}{i} 10830~\AA}~were made in parallel
with the Tenerife Infrared Polarimeter \citep[\mbox{TIP-II,}][]{Collados2007}
and the full-disk Chromospheric Telescope
\citep[ChroTel,][]{Kentischer2008,Bethge2011} at the German
Vacuum Tower telescope, Iza\~na, Spain. The same observations were
used before to calibrate the derivation of velocities from ChroTel
intensity filtergrams by a comparison to simultaneous high-resolution spectra (see
\citealt{Bethge2011} for a description of the approach). Simultaneous
with the He spectra from TIP-II (just termed `TIP' in the following), we used the POlarimetric LIttrow
Spectrograph \citep[POLIS,][]{beck+etal2005} to obtain co-spatial
intensity spectra of \mbox{\ion{Ca}{ii} H}. 
\begin{figure*}
 \centering
 \includegraphics[width=0.825\textwidth]{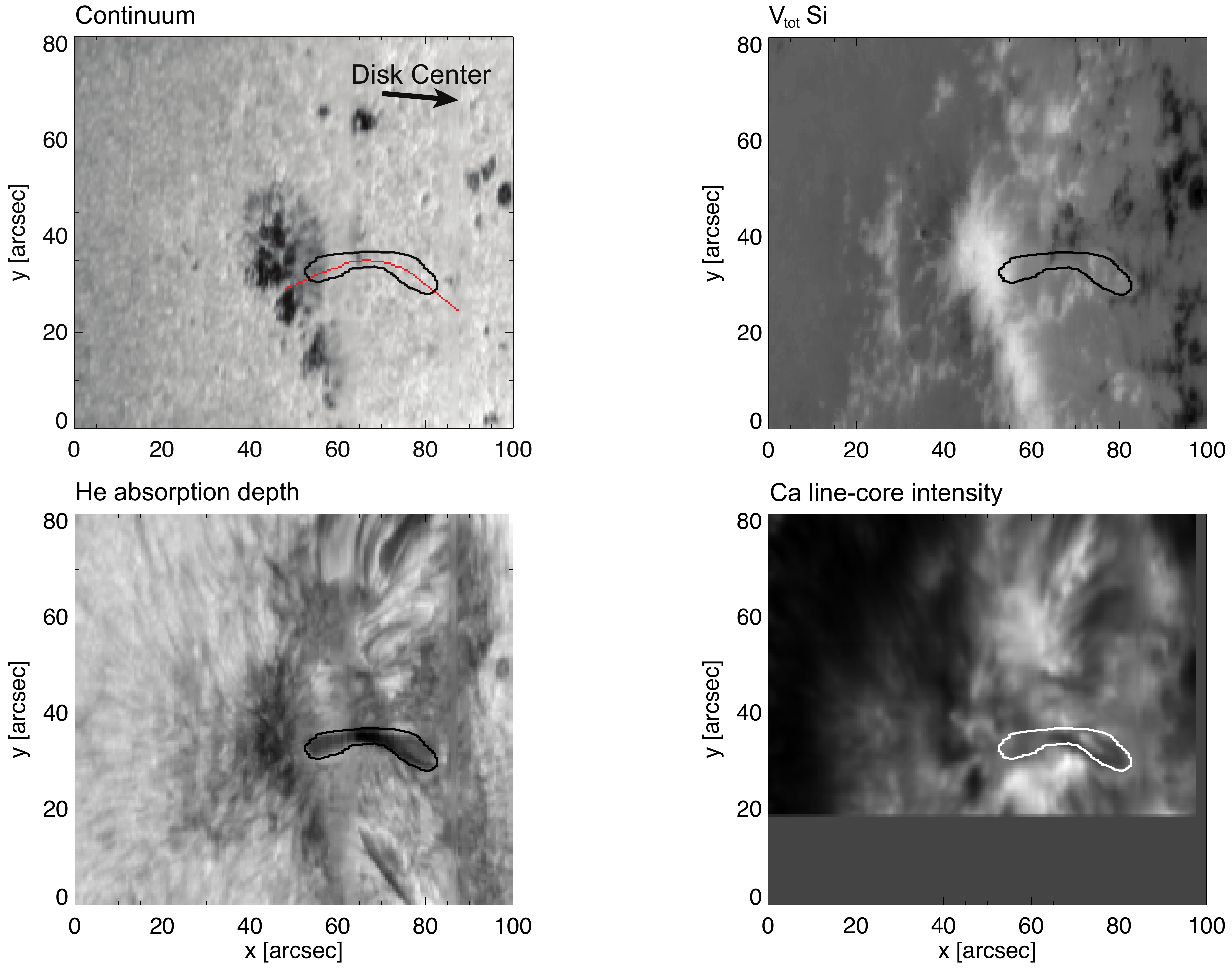}
  \caption{Overview maps of the observed FOV. \emph{Clockwise from
      upper left:} continuum intensity in the infrared, total circular
    polarization of \mbox{\ion{Si}{i}}, line-core intensity of \mbox{\ion{Ca}{ii}
    H}, absorption depth of \mbox{\ion{He}{i}}. The contour outlines the
    structure hosting the siphon flow under investigation.
    All maps are shown on a linear scale, the total circular polarization
    is scaled over a range of [$-$3\,\%, 6\,\%] of the continuum
    intensity. The red line in the continuum map denotes the central
    spine of the structure.    
    \label{fig:siphon_tip}} 
\end{figure*}

\begin{table}
\caption{Line parameters.\label{tab:tab_lines}}
\begin{tabular}{lllll}\hline\hline
No. & Ion  & $\lambdaup$ [nm] & height umbra [km] & height QS [km]  \cr\hline\\[-8.5pt]
1 & \ion{Fe}{i}  & 396.55  & 558\tablefootmark{3} &  $\sim$250\tablefootmark{1} \cr
2 & \ion{Fe}{i}  & 396.61 &  568\tablefootmark{3} & 450\tablefootmark{1} \cr
3 & \ion{Fe}{i}  & 396.93  & 588\tablefootmark{3} & 600\tablefootmark{1} \cr\hline\\[-8.5pt]
4 & \ion{Ca}{ii} & 396.85 &   $\sim$1000\tablefootmark{3} & $\sim$1000\tablefootmark{1} \cr
5 & \ion{Si}{i}  & 1082.71 & 300\tablefootmark{2,3} & 540\tablefootmark{2} \cr
6 & \ion{He}{i}  & 1083.03 & $>$1000\tablefootmark{3} & -- \cr\hline\\[-7pt]
\end{tabular}
{\bfseries Notes.}
Line formation heights in the umbra and quiet Sun (QS) are taken from
\tablefoottext{1}{\citet{beck+etal2009}},
\tablefoottext{2}{\citet{bard+carlsson2008}}, and
\tablefoottext{3}{\citet{felipe+etal2010}}.
\end{table}
\subsection{Spectropolarimetric and spectroscopic data}
The spatial scan with the two slit-spectrographs TIP and POLIS
covered a part of the active region NOAA 10978 on 8 December 2007 at
68$^{\circ}$\,E and 14.5$^{\circ}$\,S, i.e., at \mbox{$\cos\thetaup=$
  0.63}. The scanning procedure took 15 minutes from 11:05--11:20
UT. The total size of the map was 100\arcsec$\times$82\arcsec\ for TIP and
100\arcsec$\times$70\arcsec\ for POLIS, consisting of 200 steps of
0\farcs5 step width. The spatial sampling along the slit was 0\farcs3
for Ca and 0\farcs18 for TIP, which was binned by two to 0\farcs36
afterwards. The slit width was 0\farcs36 for TIP and 0\farcs5 for
POLIS, respectively. The spectral resolution of the TIP (POLIS)
data was 10.91 (19.2)~m\AA/pixel, the spatial resolution is estimated
to be about 1\arcsec. We used an integration time of 3 seconds per
scan step. The mean noise level of the TIP spectra in Stokes $I$
and Stokes $V$ is 1.5$\,\cdot\,$10$^{-3}$ and 3.9$\,\cdot\,$10$^{-3}$
times the continuum intensity, respectively. The TIP data
cover a spectral range of 10823.25--\mbox{10834.25~\AA}, so both the
photospheric \mbox{\ion{Si}{i}} 10827~\AA$\;$line and the chromospheric
\mbox{\ion{He}{i}} 10830~\AA$\;$triplet are contained in the spectra
(shortened `\mbox{\ion{Si}{i}}' and `\mbox{\ion{He}{i}}' in the
following). The \mbox{\ion{Ca}{ii} H} (`\mbox{\ion{Ca}{ii}}' in the following)
spectra from POLIS cover the core and the blue wing of the line that
contains several photospheric blends. Table~\ref{tab:tab_lines} lists the
observed spectral lines, together with estimates of their formation
height in the quiet Sun (QS) and in the umbra of sunspots. 

The total linear and total circular polarization (L$_{\rm{tot}}$ and
V$_{\rm{tot}}$) were calculated separately for \mbox{\ion{Si}{i}} and
\mbox{\ion{He}{i}} according to the definition given in \cite{Lites1999} as a
first estimate of the magnetic field configuration in the photosphere
and chromosphere.

The LOS velocity in \mbox{\ion{Si}{i}} was determined in every pixel by looking
for the position of the minimum intensity of the (smoothed) spectrum in the
wavelength range between 10826.6 and 10827.4~\AA. For the velocities
in \mbox{\ion{He}{i}}, the chosen wavelength range was between 10828.5 and
10831.9~\AA, i.e., between the outer wing of the \mbox{\ion{Si}{i}} line and
the water vapor line on the red side of the \mbox{\ion{He}{i}} line. The
minimum intensity and absorption depth refer to the stronger red component of
the \mbox{\ion{He}{i}} triplet. The determination of the line shifts was
done with pixel precision in the spectra, corresponding to an uncertainty in
velocity of about 300~m~s$^{-1}$. For the \mbox{\ion{Ca}{ii}} spectra from POLIS, we
determined the LOS velocities of three of the photospheric line blends
and of the very line core. For locations with a single emission peak
in \mbox{\ion{Ca}{ii}} instead of an absorption core such as found in the umbra, we
calculated the center-of-gravity (COG) of the emission pattern. This
quantity yields a similar spatial distribution of velocities across
the FOV, but with a reduced dynamic range, because the absorption core
is only one component of the central emission. We thus scaled the COG
velocities up by a factor of about 10, which was motivated through a
scatterplot between COG velocities and line-core velocities. We then
substituted the unreliable line-core velocity values on the locations
of single emission peaks \citep[cf.][]{rezaei+etal2008,beck+etal2009}
with the corresponding values of the COG. 

An overview of the observed region on the Sun is shown in Fig.~\ref{fig:siphon_tip}.
To mark the relevant structure for the present investigation, we
manually drew a contour line around the feature in the \mbox{\ion{He}{i}}
Doppler map ({\em top right panel} of Fig.~\ref{fig:velo_plot}).
The latter is overlayed on all other maps, it also coincides with the
boundary of the structure in the He absorption depth and the \mbox{\ion{Ca}{ii}} core
intensity (see Fig.~\ref{fig:siphon_tip}). 

\subsection{Chromospheric Doppler shift imaging: ChroTel}
ChroTel full-disk filtergrams in \mbox{\ion{He}{i}} 10830~\AA$\;$were acquired
before, during, and after the spectrograph scan with a cadence of
30~s. Data are available from 9:07 to 12:19 UT with some gaps
caused by the flatfielding procedure. The same region that was scanned
with TIP and POLIS was cut out from the ChroTel filtergrams
and carefully aligned with the spectrograph data.  

\begin{figure*}
 \centering
 \includegraphics[width=0.829\textwidth]{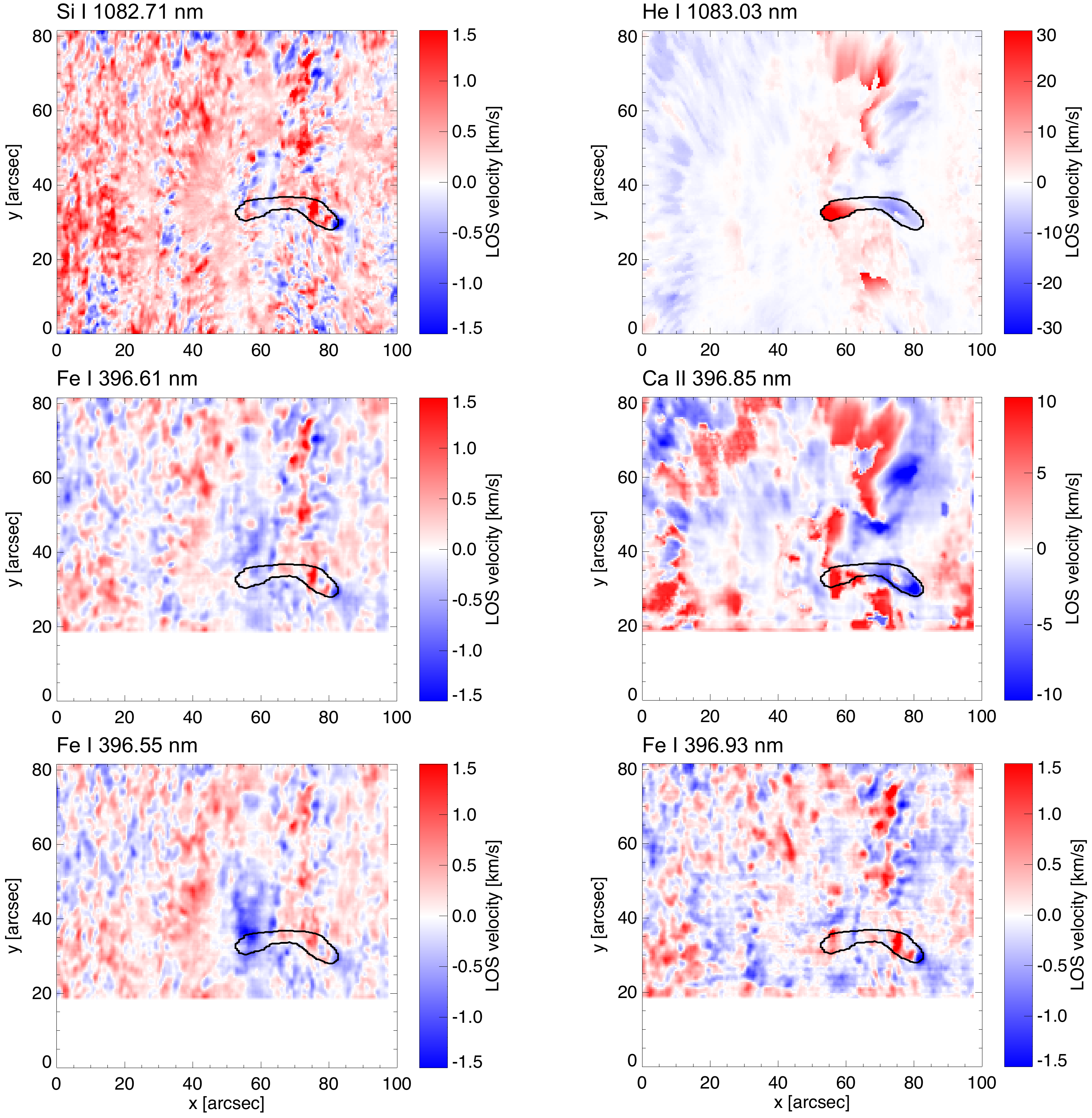}
\caption{Line-of-sight velocities. {\em Left column, bottom to top}: \mbox{\ion{Fe}{i}}
  at 396.55~nm, \mbox{\ion{Fe}{i}} at 396.61~nm, \mbox{\ion{Si}{i}} at
  1082.71~nm. {\em Right column, bottom to top}: \mbox{\ion{Fe}{i}} at
  396.93~nm, \mbox{\ion{Ca}{ii} H} core at 396.85~nm,
  \mbox{\ion{He}{i}} at 1083.03~nm. \label{fig:velo_plot}} 
\end{figure*}

With a tunable Lyot-type filter, ChroTel takes seven filtergrams
(FWHM\,=\,1.3~\AA) in and around the \mbox{\ion{He}{i}} triplet within less
than 8~s. As shown in \cite{Bethge2011}, it is possible to derive line
shifts and therefore LOS velocities from these
filtergrams. \mbox{\ion{He}{i}} absorption depth maps and LOS velocity maps
were created for the region of interest about two hours before,
fifteen minutes before, during, and one hour after the TIP scan. To
derive these maps, the data were averaged over 7~minutes to increase the
S/N ratio. \cite{Bethge2011} showed that high velocities 
($>\,${\raise.17ex\hbox{$\scriptstyle\sim$}}5~km~s$^{-1}$) are
underestimated by a factor of about 3 in the ChroTel maps. The color
bars for the Doppler velocities in Fig.~\ref{fig:temporal} are scaled
accordingly for an easier comparison with the LOS velocities shown in
Fig.~\ref{fig:velo_plot}.   

\subsection{Coronal context data: TRACE}
Raw \mbox{\ion{Fe}{ix/x}} 171~\AA\,data from the Transition Region
And Coronal Explorer \citep[TRACE,][]{Handy1999}, co-temporal with the
ChroTel data, are employed for coronal context. The emission in
this band samples plasma at a temperature of about 10$^6$~K. 

The TRACE maps were rotated, rescaled, and aligned. The
alignment was done by eye, both with the ChroTel \mbox{\ion{He}{i}} maps and among
the TRACE maps themselves. Since \mbox{\ion{Fe}{ix/x}} 171~\AA$\;$and \mbox{\ion{He}{i}}
10830~\AA$\;$sample completely different layers, of course not all
features in the TRACE data are commensurate to features seen with
ChroTel. Nevertheless, the observed region exhibits quite
characteristic structures and we estimate the alignment to be no
worse than 2\arcsec\ to 3\arcsec. 

\section{Hydrodynamics of the flow\label{sec:hotf}}
\subsection{Morphology and dynamics at high spatial resolution}
The structure outlined in Fig.~\ref{fig:siphon_tip} has a (projected)
size of about 30\arcsec\,$\times$\,6\arcsec, i.e., about 21~Mm\,$\times$\,4~Mm on the Sun. The contour was
selected by eye roughly along the strongest velocity gradients
seen in \mbox{\ion{He}{i}} outside of the apparently coherent region
of the dark loop-like feature seen in \mbox{\ion{He}{i}} (cf.~{\em lower left
  panel} of Fig.~\ref{fig:siphon_tip} and {\em top right panel} of
  Fig.~\ref{fig:velo_plot}). The \mbox{\ion{He}{i}} absorption pattern coincides
  well with the outlined 
structure based on the velocities. The same holds for the line-core 
intensity of \mbox{\ion{Ca}{ii}} that, however, also shows more substructure with a
brightening in the middle towards the right end of the structure. The
\mbox{\ion{He}{i}} absorption is strongest in the middle of the contour where
the LOS velocities are close to zero.    

The structure terminates at its left end (`inner footpoint' in the
following) in the penumbra of a sunspot with positive polarity in the
map of the signed total circular polarization
($\textrm{V}_{\textrm{\tiny tot}}$) of the photospheric \mbox{\ion{Si}{i}}
line (Fig.~\ref{fig:siphon_tip}, {\em top right}). At the right end (`outer footpoint'), black patches are seen
in the same map just outside of the contour, indicating the opposite
direction of the magnetic field than in the penumbra. The LOS
velocities of both chromospheric lines (\mbox{\ion{He}{i}} and
\mbox{\ion{Ca}{ii} H},
Fig.~\ref{fig:velo_plot}) show a smooth change from upflows 
at the outer footpoint in the QS to downflows of up to 40\,km~s$^{-1}$
in the penumbra. The surroundings of the feature show no strong
upflows outside the contour in the two chromospheric lines, especially
none to the right of the outer footpoint. The same holds for the
downflows at the inner footpoint: they extend as far as the contour
line into the penumbra, but disappear abruptly there.  

The various photospheric lines ({\em left column} and {\em lower right
  column} of Fig.~\ref{fig:velo_plot}) have been roughly sorted by their
corresponding formation height, from bottom up in each column. They
show a clear pattern: in the lowermost forming lines ({\em left
  column}), the inner footpoint with downflows is not clearly seen
because the lines are dominated by the line shift from the Evershed
effect, yielding upflows on the center-side penumbra and downflows on
the limb-side penumbra. The signature of the Evershed flow diminishes with
increasing formation height, and reveals a patch of downflows near the
inner end of the contour in the higher atmosphere. These downflows are
most pronounced and cover the largest area in the \mbox{\ion{Fe}{i}} line at
396.93\,nm ({\em lowermost panel} in the {\em right column}), but can
also be seen for \mbox{\ion{Si}{i}} ({\em topmost panel} in the {\em left
  column}) and rudimentarily in the \mbox{\ion{Fe}{i}} line at
396.61\,nm ({\em middle panel} in the {\em left column}). In the latter two lines, the
downflows are just at the edge of the contour. For the outer
footpoint, the case is clearer. All photospheric lines show upflows at
the very end and outside of the contour, whose amplitude increases
with formation height from 0.5\,km~s$^{-1}$ to 1\,km~s$^{-1}$ in,
e.g., \mbox{\ion{Si}{i}}. The upflows are more localized than the downflows
and only slightly increase in area in the upper atmosphere. 

\begin{figure}
\resizebox{8.8cm}{!}{\includegraphics{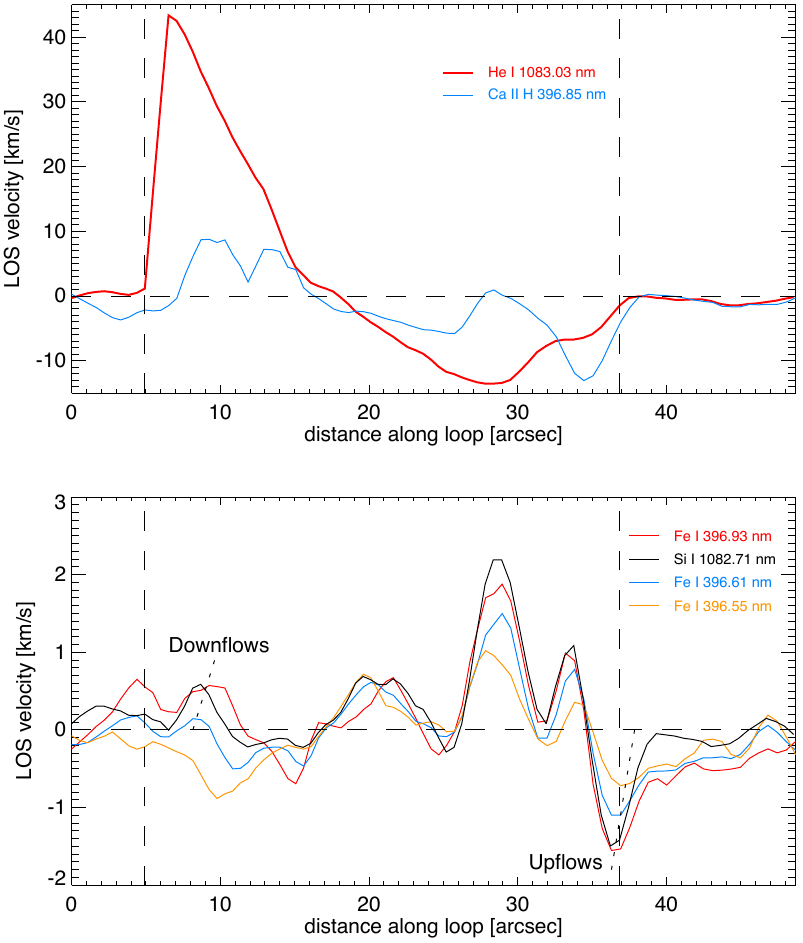}}
\caption{Line-of-sight velocities along the central axis of the loop
  structure under investigation. {\em Top panel}: 
  chromospheric lines of \mbox{\ion{He}{i}} and \mbox{\ion{Ca}{ii} H}. 
  {\em Bottom panel}: photospheric lines.
  The {\em dashed vertical lines} denote the inner and outer
  end of the structure. The {\em dotted lines} indicate the locations
  of upflows and downflows.
  \label{fig:velo_along}} 
\end{figure}

Within the structure, the \mbox{\ion{He}{i}} velocities are in the range from
$-$6$\;$to$\;-$12~km~s$^{-1}$ in the upflow region and up to +43~km~s$^{-1}$ at
the far end of the downflow region. In between, the velocities gradually 
change from blueshifts to redshifts, with a slight tilt of the
isovelocity levels relative to the spine of the structure. The
velocity structure is visualized in the LOS velocities along the
central axis of the structure, which are shown in Fig.~\ref{fig:velo_along}. The
photospheric lines ({\em lower panel}) show upflows at the outer
footpoint of about 1$-$2$^{\prime\prime}$ extent, and more
spatially extended downflow patches near the inner footpoint. There
the velocity reduces with formation height, with the maximum downflows
located closer to the end of the structure the lower the spectral line
forms. The two chromospheric lines show the gradual change from up- to
downflows, but the trend is interrupted partly at
$d\sim28\arcsec$ near the outer footpoint. At the inner
footpoint, the velocity in, e.g., \mbox{\ion{He}{i}} drops from above
40~km~s$^{-1}$ to zero over less than 2$\arcsec$.  

\begin{figure}
\centering
\resizebox{0.985\hsize}{!}{\includegraphics{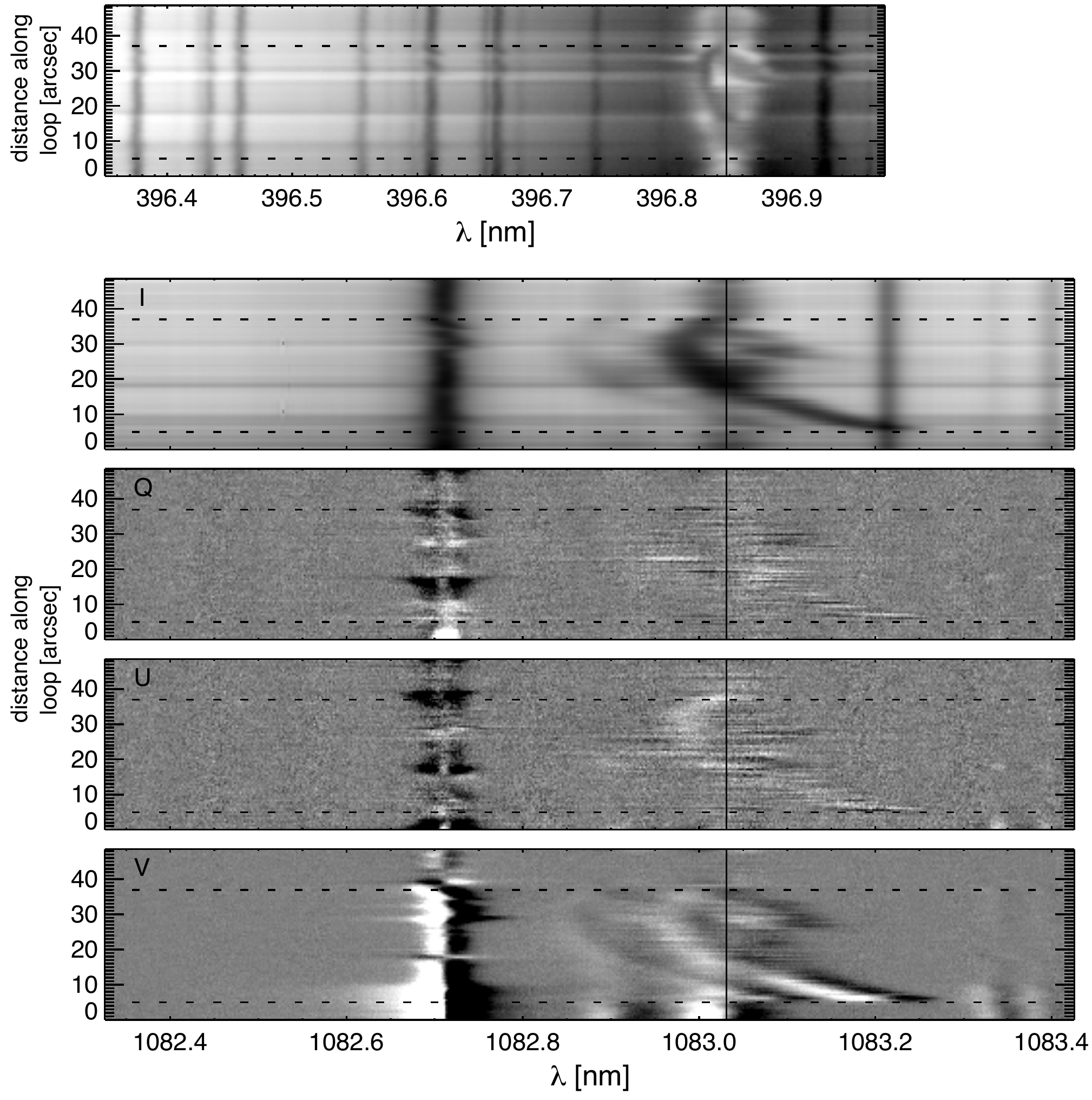}}\\
\caption{Spectra along the central axis of the structure. {\em Top to
    bottom}: intensity spectra of \mbox{\ion{Ca}{ii} H}, Stokes $IQUV$ of
  \mbox{\ion{He}{i}}. The two {\em horizontal dashed lines} in each panel
  denote the inner ({\em lower line}) and outer end ({\em upper line})
  of the structure. The {\em vertical solid lines} denote  the rest
  wavelength of the respective line.\label{fig:spectra}} 
\end{figure}

\begin{figure}[h!]
\centering
\resizebox{8.77cm}{!}{\includegraphics{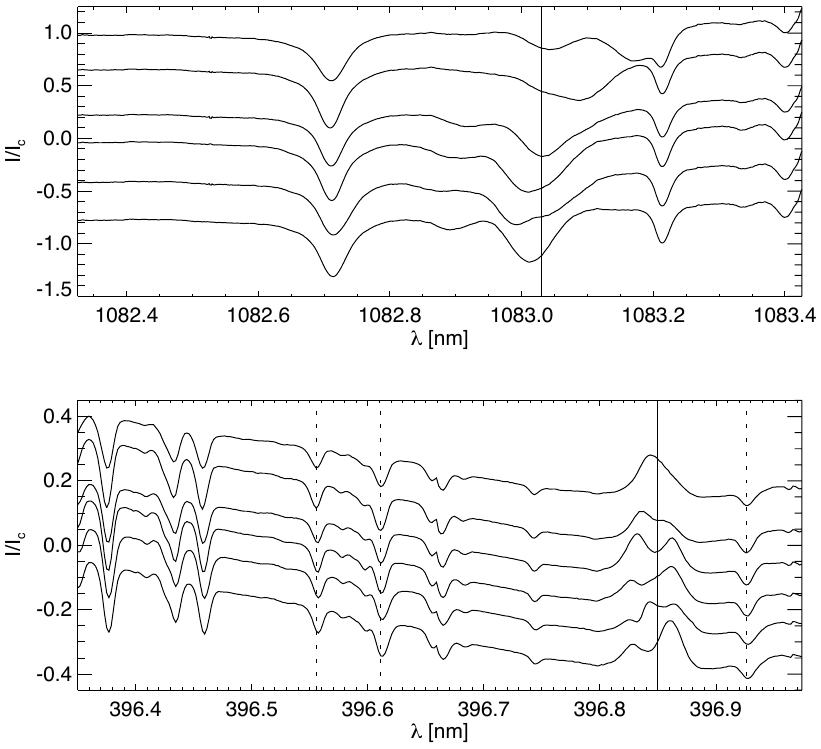}}
\caption{Individual profiles along the central axis of the structure. {\em Top/bottom
    panel}: \mbox{\ion{He}{i}}/\mbox{\ion{Ca}{ii} H}. The inner footpoint
  corresponds to the uppermost profile, the outer to the lowermost
  one. The {\em solid vertical lines} denote the rest wavelength of
  the respective line. The {\em dashed vertical lines} in the lower
  panel mark the used photospheric iron lines. The spatial separation
  between two subsequent profiles was 5$^{\prime\prime}$. The profiles were offset in
  intensity from each other arbitrarily for better
  visibility.\label{fig:profile_examples}} 
\end{figure} 

\begin{figure*}
\centering
 \resizebox{15.8cm}{!}{\includegraphics{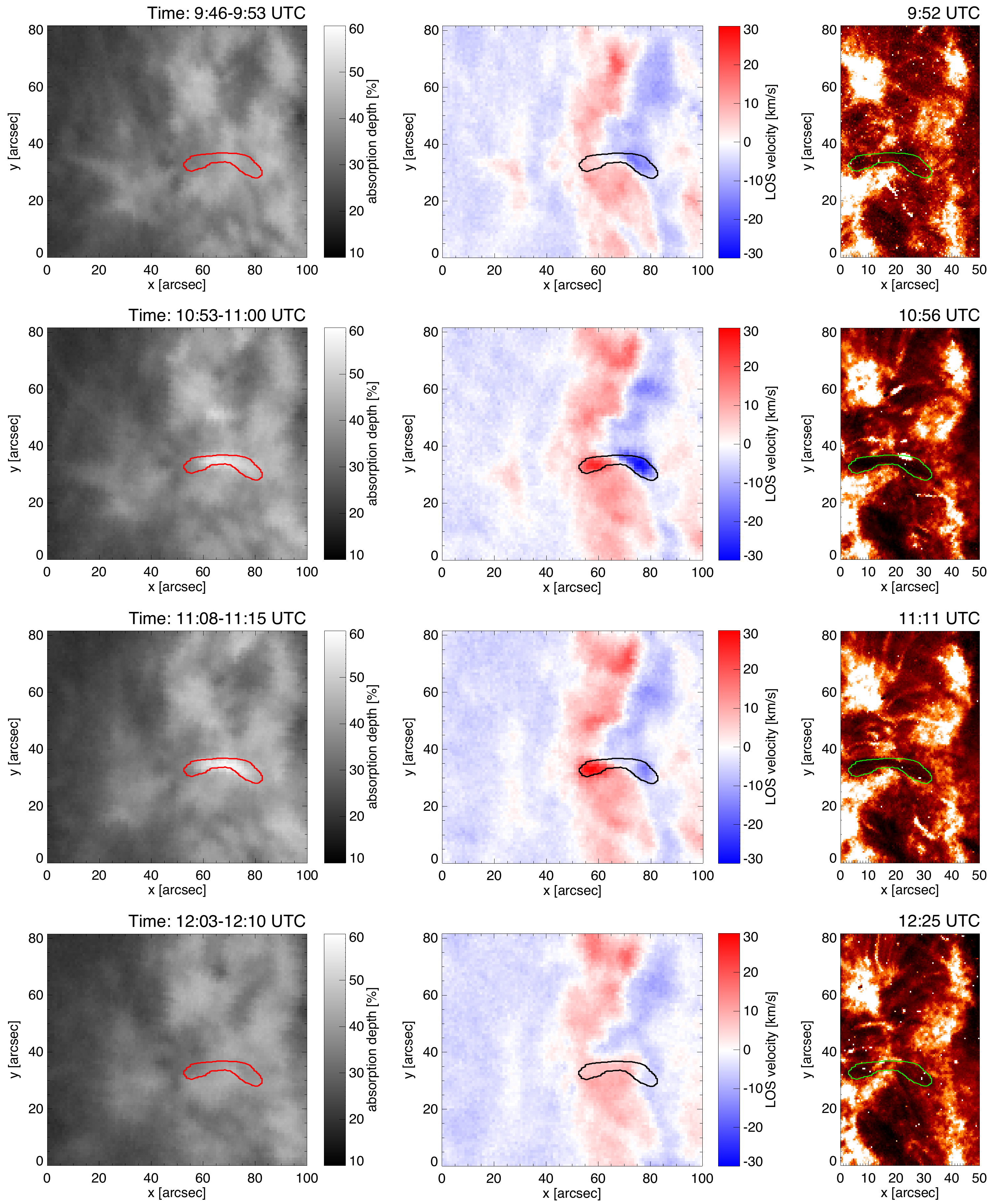}}
  \caption{Temporal evolution of the flow. \emph{Left and middle
      column:} \mbox{\ion{He}{i}} absorption depth and line-of-sight velocity from averaged
    ChroTel data. The time range denotes the averaging period for the
    data. Note that the color bars for the velocities were scaled by a
    factor of 3 to take into account the underestimation of high
    velocities in the velocity determination with the ChroTel
    filtergrams \citep[see][]{Bethge2011}. \emph{Right 
      column:} Intensity in the \mbox{\ion{Fe}{ix/x}}
    171~\AA$\;$channel of co-temporal raw TRACE data (linear scale;
    only the right half of the map is shown).}  
  \label{fig:temporal}
\end{figure*}%

The variation of the velocity trend near the outer footpoint is caused
by the presence of multiple atmospheric components that are seen in
the spectra along the central axis of the structure
(see Fig.~\ref{fig:spectra}). Whereas the structure and its corresponding
velocities can clearly be seen in the \mbox{\ion{He}{i}} and \mbox{\ion{Ca}{ii}} spectra as the
curved trace of the absorption core, additional spectral components
appear around $d\sim 28^{\prime\prime}$ with about zero velocity or
small downflows. These additional components can best be seen in the
individual \mbox{\ion{Ca}{ii}} profiles of Fig.~\ref{fig:profile_examples}. For instance,
the second \mbox{\ion{Ca}{ii}} profile from the bottom shows two different absorption
cores over a broad emission, where the structure corresponds to the one
with the large blue shift. In the \mbox{\ion{He}{i}} profiles, the components overlap
more strongly because of the intrinsic width of the absorption
components, they only can be clearly seen at the inner footpoint ({\em
  uppermost} profile): the downflows related to the structure produce
an absorption near the telluric water vapor line near 10831.5\,\AA,
whereas a second unshifted component is seen at the rest wavelength of
\mbox{\ion{He}{i}}. 

\subsection{Temporal evolution: ChroTel and TRACE\label{sec:ct_trace}}
In the ChroTel data, upflowing material is seen already two hours
before the spectrograph scan. Strong upflows with velocities of around
$-$15~km~s$^{-1}$ set in at around 9:12\,--\,9:13~UT and persist until the
first data gap at 9:23~UT. At 9:38~UT, after the data gap, the
upflows have slowed down to about $-$5 to $-$10~km~s$^{-1}$. The {\em top
row} of Fig.~\ref{fig:temporal} shows data that was averaged over the period
from 9:46\,--\,9:53~UT. Velocities around $-$6 to $-$12~km~s$^{-1}$
are seen in the upflowing branch, whereas still no enhanced downflows are
discernible yet from the general downdraft region in the
surroundings. The absorption depth in \mbox{\ion{He}{i}} shows no visible
enhancement with respect to the background.  

After another data gap of 35 minutes, the observations continued
from 10:27\,--\,11:33~UT without interruption. During this period of roughly
one hour, a continuous, but not necessarily steady flow is seen in the
structure. The second row of Fig.~\ref{fig:temporal} shows averaged data from
10:53\,--\,11:00~UT, i.e., around 15 minutes before the
spectrograph scan. The structure exhibits a symmetric flow with very
strong up- and downflow velocities of at least 30~km~s$^{-1}$. The \mbox{\ion{He}{i}}
absorption is strongly enhanced at that time, predominantly in the middle of
the structure. This is also the case in the next snapshot during the 
spectrograph scan at 11:08\,--\,11:15~UT. The velocity map confirms what
was already seen in the previous velocity maps
  (Fig.~\ref{fig:velo_plot}). After the last data gap of 30 minutes, the flow has
completely disappeared in \mbox{\ion{He}{i}} at 12:03\,--\,12:10~UT.

\begin{figure}
  \centering
  \resizebox{0.95\hsize}{!}{\includegraphics{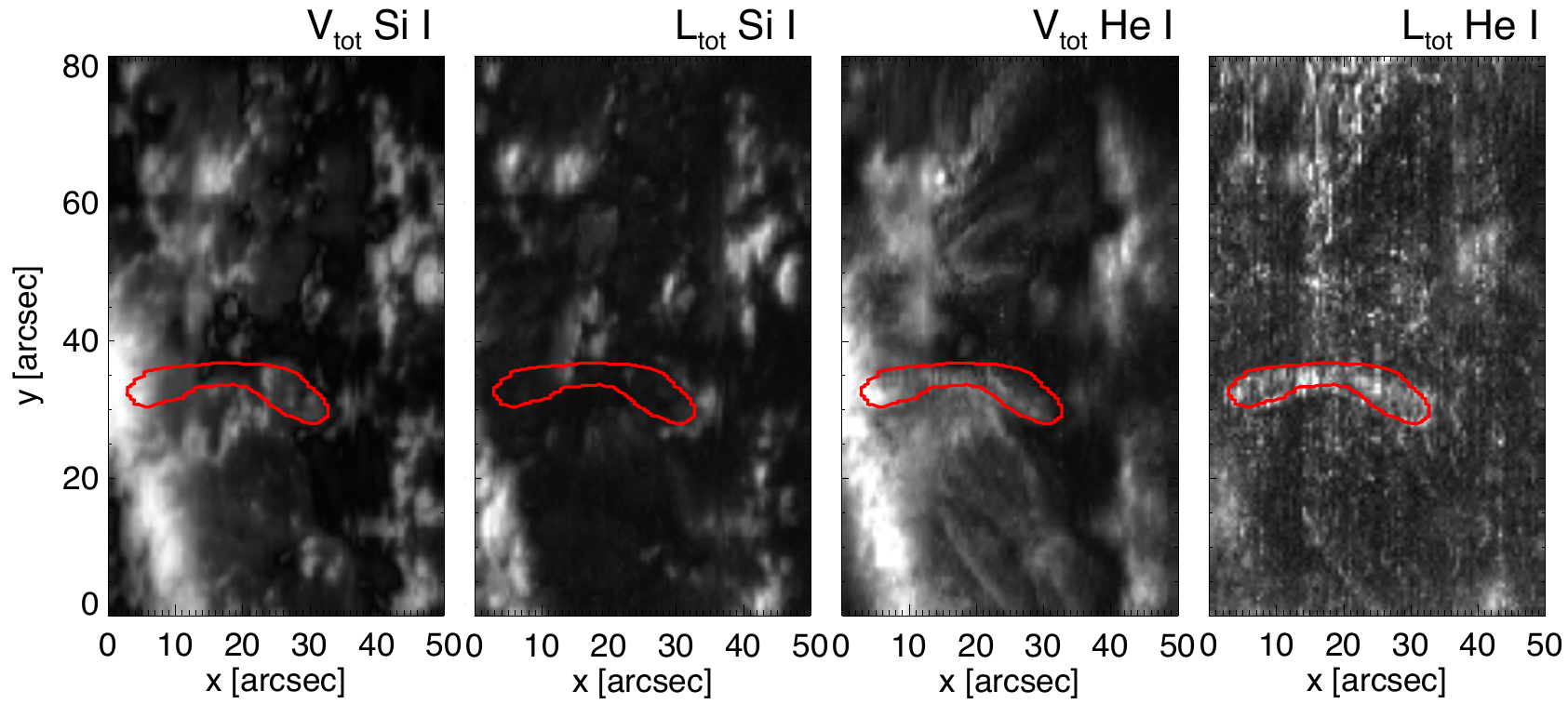}}  
  \caption{Total linear (L${_{\rm{tot}}}$) and total
    circular (V${_{\rm{tot}}}$) polarization in 
    \mbox{\ion{Si}{i}} and \mbox{\ion{He}{i}}, in percent of the
    continuum intensity. Only the right half
    of the map is shown. \emph{From left to right:}
    V${_{\rm{tot}}}$ in \mbox{\ion{Si}{i}} \mbox{(0-4~\%)},
    L${_{\rm{tot}}}$ in \mbox{\ion{Si}{i}}
    \mbox{(0-1.5~\%)}, V${_{\rm{tot}}}$ in
    \mbox{\ion{He}{i}} \mbox{(0-1~\%)}, L${_{\rm{tot}}}$ in
    \mbox{\ion{He}{i}} \mbox{(0.05-0.2~\%)}.}   
  \label{fig:polarization}
\end{figure}

In the TRACE snapshots, a dark feature is seen when also the
absorption in \mbox{\ion{He}{i}} is strong ({\em two middle rows} in
Fig.~\ref{fig:temporal}). The dark feature in the TRACE \mbox{\ion{Fe}{ix/x}} data is found exactly within the
contour derived from the \mbox{\ion{He}{i}} velocities in the TIP
  data. At 9:52\,UT, when only weak upflows were present and no
enhanced absorption in \mbox{\ion{He}{i}} is discernible yet, one gets the impression
that the so-called \emph{moss emission} from below is partially absorbed by
overlying, optically thin material in the middle of the contour. After the
flow has disappeared in \mbox{\ion{He}{i}} at 12:25 UT, the moss emission is clearly
visible again in the middle of the contour.

\section{Magnetic structure of the loop\label{sec:msotl}}

\subsection{Total polarization signals\label{sec:totp}}
Figure~\ref{fig:polarization} shows the total linear and total circular
polarization signals in \mbox{\ion{Si}{i}} and \mbox{\ion{He}{i}}. The additional
  spectropolarimetric data of the two \mbox{\ion{Fe}{i}} lines at 630\,nm from the
  second channel of POLIS have not been included in the current investigation,
  because they are similar to the \mbox{\ion{Si}{i}} line. The polarization
signals in \mbox{\ion{Si}{i}} show no exceptional match with the contour,
which is expected because only the footpoint signatures should be
visible in the photosphere. As the map of the signed total circular
  polarization in \mbox{\ion{Si}{i}} showed, patches with a polarity opposite
  to that at the downflow footpoint are seen at the very border of the
contour in the upflow region, i.e., the upflow footpoint is likely to be
located in these regions of opposite polarity.  

\begin{figure}
\centering
\centerline{\resizebox{7.5cm}{!}{\includegraphics{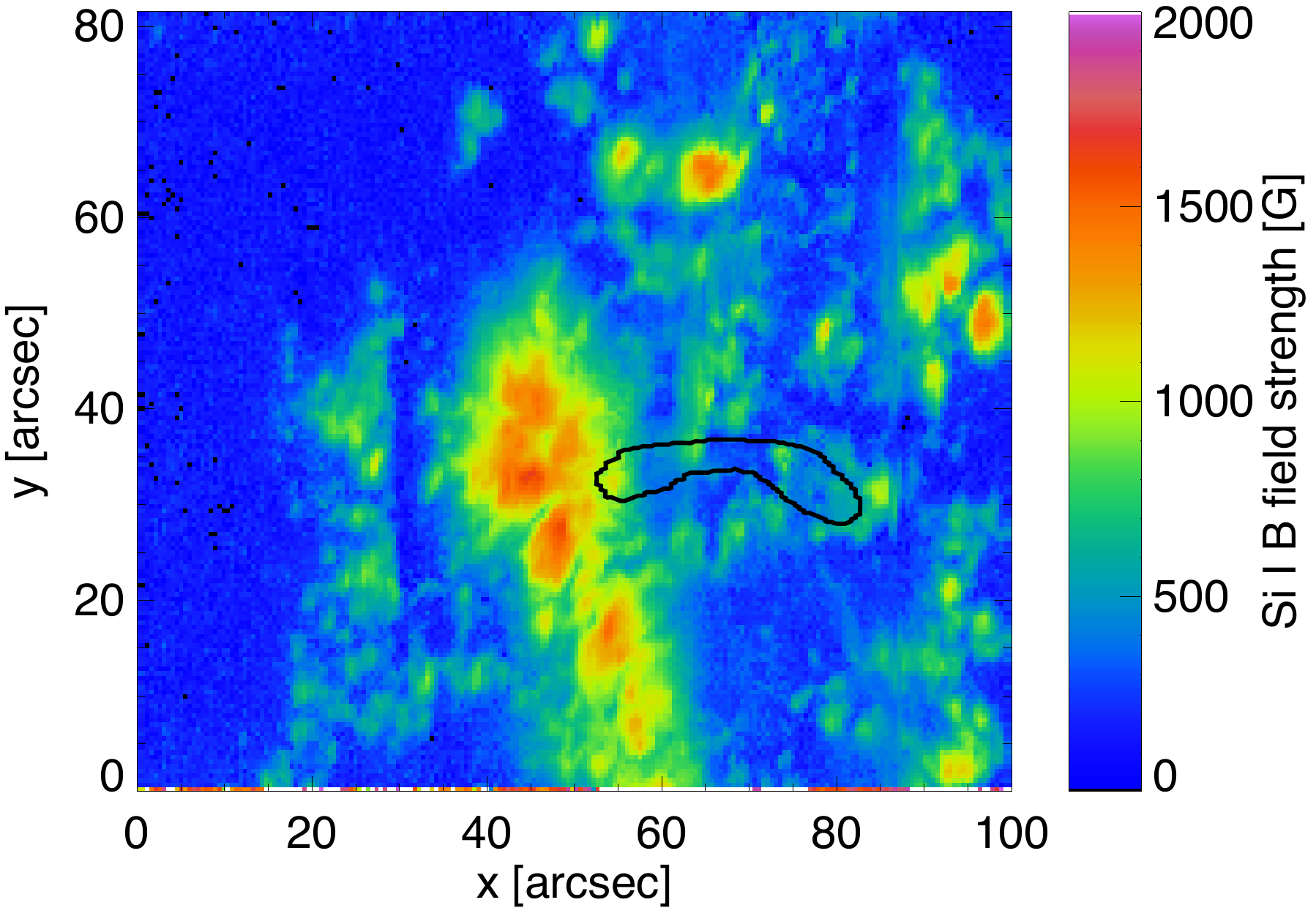}}}
\caption{Magnetic field strength in the photosphere as derived from an
  inversion of the \mbox{\ion{Si}{i}} spectra.}  
  \label{fig:inversions}
\end{figure} 

The polarization signals in \mbox{\ion{He}{i}} correspond well to the
contour. The total linear polarization signal is strongest at the location
where the velocities are close to zero. This means that we are
looking at predominantly horizontal magnetic fields and flows within,
which is likely to be the loop apex. However, both linear and circular
polarization signals are present at the same time almost everywhere in
the outlined structure, i.e., the magnetic field configuration is in
reality more complicated than the simple picture of a flux tube
arching from one footpoint to the other. The presence of multiple
  spectral components (Figs.~\ref{fig:spectra} and \ref{fig:profile_examples})
  also already indicated that the structure is not the only feature present
  in the area marked by the contour.

\subsection{Inversions of \ion{\em Si}{\em\tiny I}} \label{S:inversion.Si}
As a first estimate of the
  photospheric magnetic field properties, we inverted the spectra of the
  \mbox{\ion{Si}{i}} line with the H{\tiny
    E}LI{\tiny X}$^{+}$ Code \citep{Lagg2004,Lagg2009}. H{\tiny
    E}LI{\tiny X}$^{+}$ uses the
  Milne-Eddington approximation and provides the average values of magnetic
  field properties throughout the formation height of the line. The inversion
  of this and the other lines observed (\mbox{\ion{He}{i}}, \mbox{\ion{Fe}{i}} at 630.15\,nm
  and 630.25\,nm) will be discussed in more detail in a subsequent
  publication. Here, we only show the map of the
  photospheric magnetic field strength as derived from the
  \mbox{\ion{Si}{i}} spectra (see Fig.~\ref{fig:inversions}). In addition, we point out
  that the magnetic field inclination in the photosphere in the local reference
  frame (z perpendicular to the solar surface) from the
  inversions confirmed that the field orientation in the upflow point
  was opposite to that in the downflow point in the penumbra. 

For a comparison of the magnetic field strength in the upflow and downflow
footpoint, two regions were defined: an upflow region where the plasma is
flowing faster than $-$300~m~s$^{-1}$ in \mbox{\ion{Si}{i}}, and a downflow region
where the plasma is flowing faster than $+$35~km~s$^{-1}$ in \mbox{\ion{He}{i}}.
  The upflow region is located partly outside of the contour of the structure
  because the upflows in all photospheric lines extend slightly beyond the
  contour (Fig.~\ref{fig:velo_plot}). In both regions, the mean magnetic field
strength in the photosphere was computed from the inversion results. The
mean magnetic field strength in the upflow region is
\mbox{$\,\overline{\textrm{B}}_{\textrm{\tiny up}}=580~\textrm{G}$}, while we
find \mbox{$\,\overline{\textrm{B}}_{\textrm{\tiny down}}=950~\textrm{G}\,$}
in the downflow region. The inversions in \mbox{\ion{Si}{i}} therefore show
that the magnetic field at the inner footpoint in the sunspot
  penumbra is stronger than at the outer footpoint in the plage
  region.  

\section{Discussion\label{sec:discussion}}

\subsection{Magnitude of the siphon flow}
Driving a siphon flow through a flux tube requires a difference in gas
pressure between the two footpoints. The observations show
  that the main requirement of Eq.~(\ref{eq_siphon}) is fulfilled: the
  downflow footpoint is located in the penumbra of a sunspot with a higher
  magnetic field strength than in the upflow footpoint situated in a plage
  region. The photospheric magnetic field in the upflow region is on average
  about 1/3 weaker than in the downflow region. The total pressure, however,
  is not directly provided by the data, and could be different in both
  locations because of their fairly different surroundings. But exactly these
  surroundings provide also another argument in favor of a gas pressure
  difference: the continuum intensity, and hence, the temperature and the
  corresponding internal gas pressure, is significantly lower at the inner
  footpoint. Thus, one can reasonably assume a lower gas pressure at the inner
  footpoint that suffices to drive a siphon flow between the two points,
  flowing from weaker to stronger magnetic fields and likewise from a hotter to
  a colder region.

One can roughly estimate the magnitude of the flow based on the
difference in magnetic field. Assuming that the total \emph{external}
pressure at both footpoints is the same, the difference in gas pressure
\emph{inside} the flux tube equals the difference in magnetic
pressure. The difference in magnetic field strength of ${\Delta}B=\,$370~G at the
footpoints as mentioned in Sect.~\ref{S:inversion.Si} translates into
a magnetic pressure difference of %
\begin{equation}\label{E:delta.p}
{\Delta}p_m = \frac{({\Delta}B)^2}{2\,\mu_0} \approx 500~\mbox{J~m}^{-3}~.
\end{equation}
If we approximate the pressure gradient 
\begin{equation}\label{E:grad.p}
{\nabla}p \approx \frac{{\Delta}p_m}{L}
\end{equation}
through the pressure difference ${\Delta}p_m$ and the length of the
loop of approximately $L{\,\approx\,}20$~Mm, we can estimate the
acceleration $a$ caused by this pressure difference to be 
\begin{equation}\label{E:acceleration}
\rho\,a = {\nabla}p    
\qquad \leadsto \qquad  
a \approx \frac{{\Delta}p_m}{\rho\,L} \approx 500~\mbox{m\,s}^{-2} ~.
\end{equation}
As we estimate the effect on the chromospheric plasma seen in
\mbox{\ion{He}{i}}, we used $\rho{\,\approx\,}5{\cdot}10^{-8}$~kg~m$^{-3}$ as an estimate for the density
\citep[at a height of about 1000~km, e.g.][]{Vernazza+al:1981}. The resulting acceleration
of some ~500~m~s$^{-2}$ is about twice the solar surface
gravitational acceleration of ~$g_\odot{\,\approx\,}274$~m~s$^{-2}$, and
thus the material can be lifted up against gravity to flow along the
loop. 

The speed $v$ of the siphon flow can be roughly estimated if we assume
that the pressure difference ${\Delta}p_m$ from Eq.~(\ref{E:delta.p}),
which essentially is a (magnetic) energy density, is turned into
kinetic energy:
\begin{equation}
\frac{1}{2}\,\rho\,v^2 = {\Delta}p_m ~.
\end{equation}
Here, we also assume that the two footpoints are at the same height, i.e.,
that they have the same potential energy. This yields a velocity of
about $v\approx140\,\mbox{km~s}^{-1}$.  

The observations show speeds along the line-of-sight of more than
40~km~s$^{-1}$ (see Fig.~\ref{fig:velo_along}). Depending on the loop
geometry, this is a lower limit for the true velocity along the loop. 
From this we can conclude that the speed of
about $140~\mbox{km~s}^{-1}$ that we have roughly estimated for the
siphon flow is consistent with the observed flows. 

\subsection{Comparison to siphon flow models}

During the whole observation, the moving plasma stays confined within the
outlined structure, as it becomes apparent from the ChroTel data
and the absorption features seen in the TRACE images (see
Sect.~\ref{sec:ct_trace}). The total polarization signals in
\mbox{\ion{He}{i}} coincide extremely well with the structure outlined from
the velocities (Sect.~\ref{sec:totp}), supporting the view that the plasma is
moving within a magnetic loop in the chromosphere. 

The ChroTel data show a continuous flow for at least one hour; the
direction of the flow is always maintained. \cite{Cargill1980} found that pressure perturbations in already
existing flows can lead to reverse flows from the footpoint with
lower gas pressure to the one with higher gas pressure. In this
observation, this seems not to be the case. Nevertheless, the
unidirectional flow observed here goes through different stages, from
an asymmetrical velocity distribution with little \mbox{\ion{He}{i}}
absorption and subsonic flow speeds to a very symmetrical flow with
strong \mbox{\ion{He}{i}} absorption and velocities that clearly exceed the
sound speed of about 10~km~s$^{-1}$ at typical formation temperatures
of the \mbox{\ion{He}{i}} triplet. At first sight, the latter
stage seems to resemble the `purely supercritical flow' steady-state
solutions for siphon flows described in \cite{Cargill1980} or in
\citet[his Fig.~3\,d]{Thomas1988}. In these solutions, the velocity
increases monotonically and the gas pressure decreases monotonically
all along the descending part of the tube. In reality however, the
downflow must become subcritical at some point to fulfill the
pressure balance condition at (or beneath) the photospheric downstream
footpoint. Because the \ion{He}{i} observations are limited to
chromospheric heights, they should therefore rather represent
something similar to the upper parts of the flux tube in the `critical
solution' of \citet[his Fig.~3\,c]{Thomas1988}. In this solution, the flow speeds
are only supercritical near the loop apex and become subcritical further down
in the atmosphere.

The velocity distribution during the spectrograph scan is completely compatible
with the `critical solutions' in both \cite{Cargill1980} and
\citet[his Fig.~3\,c]{Thomas1988}, with subsonic upflow velocities  and
supersonic downflow velocities (Figs.~\ref{fig:velo_plot} and
  \ref{fig:velo_along}). For this case, a tube shock in the downflow branch is
predicted, where the velocities undergo a sharp transition from super- to
subsonic. In \mbox{\ion{He}{i}}, there are indeed two velocity components present in
the spectra in most pixels at the edge of the downflowing branch: a fast
component at about 40~km~s$^{-1}$ and a slow component at about
5~km~s$^{-1}$. However, the latter might also represent plasma from the
general downdraft region in the background that is consistently present during
the observations.

 In the solution of \cite{Thomas1988}, the sharp
  velocity transition is accompanied by an increase in tube diameter
  and a decrease in magnetic field strength. Whereas for the latter we find no
  direct indications, the former is supported by the highest-forming
  photospheric spectral line \mbox{\ion{Fe}{i}} at 396.93\,nm. Even if it already
  shows only subsonic velocities, the size of the downflow patch is
  clearly larger than the very localized upflow patch
  (Fig.~\ref{fig:velo_along}). The fact that only small redshifts are seen
  near the inner footpoint in all photospheric lines means that the
  deceleration of the plasma must have taken place between the 
  formation heights of \mbox{\ion{He}{i}} or \mbox{\ion{Ca}{ii} H} and the upper photosphere
  (about 600\,km height, cf.~Table \ref{tab:tab_lines}). The tube shock -- if
  present -- has therefore occurred in an intermediate layer, where the plasma
  is already too cool to be visible in \mbox{\ion{He}{i}}, but still too hot to
  appear in the photospheric lines. The emission near the \mbox{\ion{Ca}{ii}} line core
  presents the only evidence in favor of a possible tube shock. At the inner
  end of the structure, the core emission is slightly enhanced (at the {\em
    lower dashed horizontal line} in the {\em upper panel} of
  Fig.~\ref{fig:spectra}). If the tube shock has thus occurred in an
  intermediate layer, it is of course not surprising that no observational
  signature is present in \mbox{\ion{He}{i}} or photospheric lines. Also, the
  theoretical calculations referred to a \mbox{high-$\betaup$} plasma, i.e.,
  for conditions found primarily in the photosphere. Since the sharp velocity
  transition occurs in a layer above the formation height of \mbox{\ion{Fe}{i}} at
  396.93\,nm, the condition of a \mbox{high--$\betaup$} plasma might not
  apply. This is supported by the fact that the flows seen in \mbox{\ion{He}{i}} are
  confined within the outlined structure at all times, suggesting that within
  the structure the magnetic field dominates the plasma motions at any height.

\subsection{Relation to TRACE 171\,\AA\ observations} \label{S:TRACE}

That the dark structures in the TRACE 171~{\AA} images represent the same
material that is seen in \mbox{\ion{He}{i}} 10830~\AA$\;$is suggested not
only by the remarkable correlation with the contour, but also from a
study from \cite{Anzer2005}. They showed that the \mbox{\ion{Fe}{ix/x}}
radiation around 171\,\AA$\;$originating from a 10$^6$~K hot plasma
can be effectively absorbed by the Lyman continuum of neutral
hydrogen ($\lambda<912~$\AA), by neutral helium ($\lambda<504~$\AA),
and by singly ionized helium ($\lambda<228~$\AA). The cool material can
therefore absorb the moss emission originating from below the structure.

It is an
assumption that the emission under the loop is present during the whole
observation, but there are three arguments in favor: (1) The moss emission
is visible through a somewhat `semi-transparent' layer when the siphon
flow sets in. (2) It can be seen on either side of the middle of the
contour at all times. (3) It is present in the contour
after the siphon flow has disappeared. The latter is the strongest
indication for constant presence of the moss emission below the
structure. This would mean that the magnetic loop in which the siphon
flow takes place is embedded in the corona in its uppermost parts.
According to 3D MHD coronal models
\citep[e.g.,][]{Bingert+Peter:2011} the magnetic transition into the
corona occurs at some 5~Mm height. Thus one could assume that the
apex of the loop is located well above 5~Mm (while the footpoints are
separated by about 20~Mm).

This would also
explain the high absorption depth in \mbox{\ion{He}{i}} that is seen in the
middle of the structure, which is probably not only caused by the
higher plasma density inside the loop, but also by the enhanced EUV
radiation from the corona in the surroundings that populates the
triplet states of the \mbox{\ion{He}{i}} 10830\,\AA~transition
\citep{Centeno2008}. While the coronal illumination reaches the
lower-lying parts of the loop only from above and is additionally
partially absorbed on its way to the chromosphere, the coronal part of
the loop gets illuminated `in situ' and from all directions, leading
to a higher population of the triplet states and therefore to a stronger
absorption in \mbox{\ion{He}{i}}. 

In a future study we will investigate the magnetic topology of the
siphon flow in the photosphere and in the chromosphere from inversions of all
observed lines, and determine the mass flux balance through the
structure from one end to the other. 

\section{Initiation and termination of the flow\label{sec:speculation}}

\begin{figure}
\centering
  \resizebox{0.997\hsize}{!}{\includegraphics{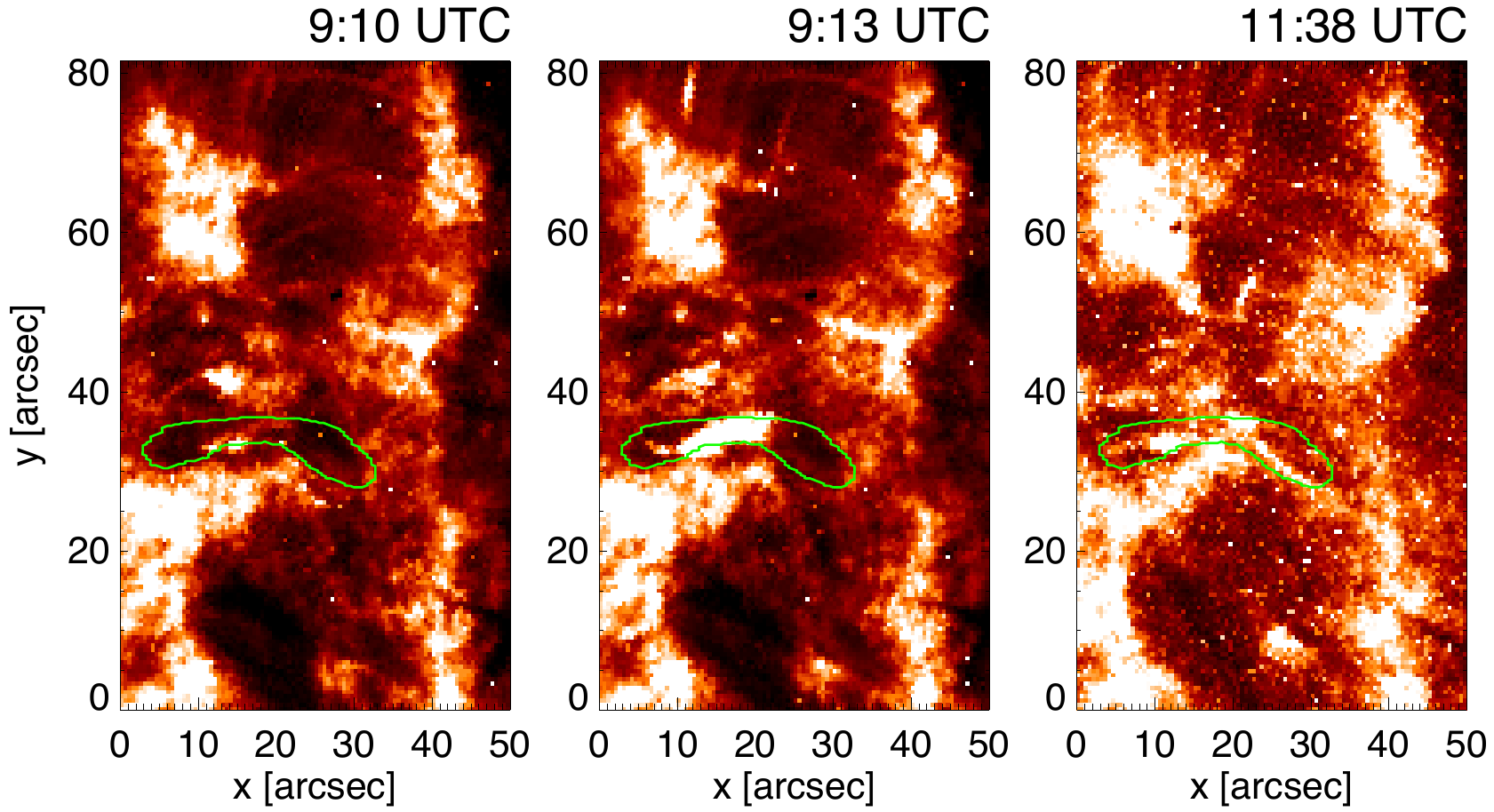}}
  \caption{The siphon flow region in TRACE \mbox{\ion{Fe}{ix/x}}
    171~\AA$\;$data when the strong upflows in \mbox{\ion{He}{i}} set in (9:13
  UT), just before that for comparison (9:10 UT), and within the data
  gap in the ChroTel data in which the flow disappeared (11:38
  UT). Notice the brightenings within the contour in the two rightmost
  images.} 
  \label{fig:brightening}
\end{figure}

In the following we will give some speculations on what might have triggered the onset and termination of the
flow. Figure \ref{fig:brightening} shows TRACE images at three
times: at 9:13~UT when the strong upflow velocities started in
\mbox{\ion{He}{i}}, at 9:10~UT just before that for comparison, and at
11:38~UT, which is the last point in time before the flow disappeared
and for which TRACE data was still available. 

At 9:13~UT, at about the same time when the strong upflows start in
\mbox{\ion{He}{i}}, a brightening appears inside the contour. It becomes clear
when the TRACE images are watched as a movie that this is not moss emission from
below, but a short-lived ({\raise.17ex\hbox{$\scriptstyle\sim$}}3~min)
brightening inside the (already existing) structure. This brightening could
be caused by heating as a result of reconnection near the apex while
the loop-like structure is presumably rising into the upper
atmosphere. Then the magnetic field lines eventually reconnected to a
different location in the (subsequent) downflow region, a location
with a magnetic field strength large enough to initiate the siphon flow because of the
difference in gas pressure between the new inner and the old outer footpoint.

Small brightenings are also seen in the contour in the last available
TRACE image at 11:38~UT, 25 minutes before the flow itself
  disappears. The idea is the same as before: a reconnection event causing
the brightening, this time leading to a disintegration of the
structure and the flow within. This is even more speculative than for
the initiation of the flow because there is no ChroTel data available at that time,
and therefore we have no information about a possible co-temporal termination of the flow. 

\section{Conclusions\label{sec:conclusions}}
We presented observational evidence for a long-lasting siphon flow of
chromospheric material in a magnetic loop structure, driven by a
difference in magnetic pressure (and therefore in gas pressure) in the
footpoints of the loop. The presented data is consistent with the
picture of cold material flowing in an arch-like structure that
reaches from the photosphere up into the corona. The flow could be
initiated and terminated by reconnecting magnetic fields, as it is
indicated from the coronal context data.

The observed difference of the magnetic field strength in the two footpoints
of the structure is sufficient to drive a siphon flow, and the
estimated speed of the siphon flow is consistent with the observed
Doppler shifts. Direct evidence of a tube shock in the downflowing branch of the loop
as predicted from theoretical work on siphon flows is not prominent in the
data. However, the weak photospheric counterpart of the strong
downflows in the chromosphere indicates that a tube shock might have
occurred at a height between about 600~km and the formation
  height of \mbox{\ion{He}{i}} ($>1000$~km). 

For the interpretation one has to follow the temporal evolution not
only in intensity images, but also in Doppler shift maps, as done here using
full-disk data from ChroTel. The line-of-sight
velocities determined from filtergram intensities in \mbox{\ion{He}{i}} are
crucial in this study to track the temporal evolution of the flow, which
emphasizes that siphon flows are dynamic phenomena going through
different phases to which existing theoretical steady-state solutions fit only
temporarily.  

\begin{acknowledgements}
\noindent The ChroTel project was funded in part by the Deutsche
Forschungsgemeinschaft (DFG). C.~Bethge would like to thank R.~Hammer,
R.~Rezaei, and R.~Schlichenmaier for valuable discussions.
  C.~Beck acknowledges support by the Spanish Ministry of Science and
  Innovation through project AYA2010--18029 and JCI-2009-04504. 
\end{acknowledgements}

\bibliographystyle{aa} 
\bibliography{literature} 

\begin{thebibliography}{30}
\expandafter\ifx\csname natexlab\endcsname\relax\def\natexlab#1{#1}\fi

\bibitem[{{Anzer} \& {Heinzel}(2005)}]{Anzer2005}
{Anzer}, U. \& {Heinzel}, P. 2005, \apj, 622, 714

\bibitem[{Aschwanden {et~al.}(2001)Aschwanden, Schrijver, \&
  Alexander}]{Aschwanden+al:2001}
Aschwanden, M.~J., Schrijver, C.~J., \& Alexander, D. 2001, \apj, 550, 1036

\bibitem[{{Bard} \& {Carlsson}(2008)}]{bard+carlsson2008}
{Bard}, S. \& {Carlsson}, M. 2008, \apj, 682, 1376

\bibitem[{{Beck} {et~al.}(2009){Beck}, {Khomenko}, {Rezaei}, \&
  {Collados}}]{beck+etal2009}
{Beck}, C., {Khomenko}, E., {Rezaei}, R., \& {Collados}, M. 2009, \aap, 507,
  453

\bibitem[{{Beck} {et~al.}(2005){Beck}, {Schmidt}, {Kentischer}, \&
  {Elmore}}]{beck+etal2005}
{Beck}, C., {Schmidt}, W., {Kentischer}, T., \& {Elmore}, D. 2005, \aap, 437,
  1159

\bibitem[{{Beck} {et~al.}(2010){Beck}, {Tritschler}, \& {W{\"o}ger}}]{Beck2010}
{Beck}, C., {Tritschler}, A., \& {W{\"o}ger}, F. 2010, Astronomische
  Nachrichten, 331, 574

\bibitem[{{Bethge} {et~al.}(2011){Bethge}, {Peter}, {Kentischer},
  {Halbgewachs}, {Elmore}, \& {Beck}}]{Bethge2011}
{Bethge}, C., {Peter}, H., {Kentischer}, T.~J., {et~al.} 2011, \aap, 534, A105

\bibitem[{Bingert \& Peter(2011)}]{Bingert+Peter:2011}
Bingert, S. \& Peter, H. 2011, \aap, 530, A112

\bibitem[{Brekke {et~al.}(1997)Brekke, Kjeldseth-Moe, \&
  Harrison}]{Brekke+al:1997:loops}
Brekke, P., Kjeldseth-Moe, O., \& Harrison, R.~A. 1997, \solphys, 175, 511

\bibitem[{{Cargill} \& {Priest}(1980)}]{Cargill1980}
{Cargill}, P.~J. \& {Priest}, E.~R. 1980, \solphys, 65, 251

\bibitem[{{Centeno} {et~al.}(2008){Centeno}, {Trujillo Bueno}, {Uitenbroek}, \&
  {Collados}}]{Centeno2008}
{Centeno}, R., {Trujillo Bueno}, J., {Uitenbroek}, H., \& {Collados}, M. 2008,
  \apj, 677, 742

\bibitem[{{Collados} {et~al.}(2007){Collados}, {Lagg}, {D{\'{\i}}az
  Garc{\'{\i}}a}, {Hern{\'a}ndez Su{\'a}rez}, {L{\'o}pez L{\'o}pez}, {P{\'a}ez
  Ma{\~n}{\'a}}, \& {Solanki}}]{Collados2007}
{Collados}, M., {Lagg}, A., {D{\'{\i}}az Garc{\'{\i}}a}, J.~J., {et~al.} 2007,
  in The Physics of Chromospheric Plasmas, ed. P.~{Heinzel}, I.~{Dorotovi{\v
  c}}, \& R.~J. {Rutten}, ASP Conference Series, 368, 611

\bibitem[{{Degenhardt}(1989)}]{Degenhardt1989}
{Degenhardt}, D. 1989, \aap, 222, 297

\bibitem[{{Degenhardt} {et~al.}(1993){Degenhardt}, {Solanki}, {Montesinos}, \&
  {Thomas}}]{Degenhardt1993}
{Degenhardt}, D., {Solanki}, S.~K., {Montesinos}, B., \& {Thomas}, J.~H. 1993,
  \aap, 279, L29

\bibitem[{{Doyle} {et~al.}(2006){Doyle}, {Taroyan}, {Ishak}, {Madjarska}, \&
  {Bradshaw}}]{Doyle2006}
{Doyle}, J.~G., {Taroyan}, Y., {Ishak}, B., {Madjarska}, M.~S., \& {Bradshaw},
  S.~J. 2006, \aap, 452, 1075

\bibitem[{{Felipe} {et~al.}(2010){Felipe}, {Khomenko}, {Collados}, \&
  {Beck}}]{felipe+etal2010}
{Felipe}, T., {Khomenko}, E., {Collados}, M., \& {Beck}, C. 2010, \apj, 722,
  131

\bibitem[{{Handy} {et~al.}(1999){Handy}, {Acton}, {Kankelborg}, {Wolfson},
  {Akin}, {Bruner}, {Caravalho}, {Catura}, {Chevalier}, {Duncan}, {Edwards},
  {Feinstein}, {Freeland}, {Friedlaender}, {Hoffmann}, {Hurlburt}, {Jurcevich},
  {Katz}, {Kelly}, {Lemen}, {Levay}, {Lindgren}, {Mathur}, {Meyer}, {Morrison},
  {Morrison}, {Nightingale}, {Pope}, {Rehse}, {Schrijver}, {Shine}, {Shing},
  {Strong}, {Tarbell}, {Title}, {Torgerson}, {Golub}, {Bookbinder}, {Caldwell},
  {Cheimets}, {Davis}, {Deluca}, {McMullen}, {Warren}, {Amato}, {Fisher},
  {Maldonado}, \& {Parkinson}}]{Handy1999}
{Handy}, B.~N., {Acton}, L.~W., {Kankelborg}, C.~C., {et~al.} 1999, \solphys,
  187, 229

\bibitem[{{Kentischer} {et~al.}(2008){Kentischer}, {Bethge}, {Elmore},
  {Friedlein}, {Halbgewachs}, {Kn{\"o}lker}, {Peter}, {Schmidt}, {Sigwarth}, \&
  {Streander}}]{Kentischer2008}
{Kentischer}, T.~J., {Bethge}, C., {Elmore}, D.~F., {et~al.} 2008, in
  Ground-based and Airborne Instrumentation for Astronomy II, Proceedings of
  SPIE 7014, 701413

\bibitem[{{Lagg} {et~al.}(2009){Lagg}, {Ishikawa}, {Merenda}, {Wiegelmann},
  {Tsuneta}, \& {Solanki}}]{Lagg2009}
{Lagg}, A., {Ishikawa}, R., {Merenda}, L., {et~al.} 2009, in The Second Hinode
  Science Meeting, ed. {B.~Lites, M.~Cheung, T.~Magara, J.~Mariska, \&
  K.~Reeves}, ASP Conference Series, 415, 327

\bibitem[{{Lagg} {et~al.}(2004){Lagg}, {Woch}, {Krupp}, \&
  {Solanki}}]{Lagg2004}
{Lagg}, A., {Woch}, J., {Krupp}, N., \& {Solanki}, S.~K. 2004, \aap, 414, 1109

\bibitem[{{Lites} {et~al.}(1999){Lites}, {Rutten}, \& {Berger}}]{Lites1999}
{Lites}, B.~W., {Rutten}, R.~J., \& {Berger}, T.~E. 1999, \apj, 517, 1013

\bibitem[{Meyer \& Schmidt(1968)}]{Meyer+Schmidt:1968}
Meyer, F. \& Schmidt, H.~U. 1968, Z. Angew. Math. Mech., 48, 218

\bibitem[{{Noci}(1981)}]{Noci1981}
{Noci}, G. 1981, \solphys, 69, 63

\bibitem[{Peres(1997)}]{Peres:1997}
Peres, G. 1997, in The corona and solar wind near minimum activity, ed.
  A.~Wilson, Proceedings of\ Fifth SOHO Workshop (ESA SP--404), 55

\bibitem[{{Rezaei} {et~al.}(2008){Rezaei}, {Bruls}, {Schmidt}, {Beck},
  {Kalkofen}, \& {Schlichenmaier}}]{rezaei+etal2008}
{Rezaei}, R., {Bruls}, J.~H.~M.~J., {Schmidt}, W., {et~al.} 2008, \aap, 484,
  503

\bibitem[{{R\"uedi} {et~al.}(1992){R\"uedi}, {Solanki}, \&
  {Rabin}}]{Rueedi1992}
{R\"uedi}, I., {Solanki}, S.~K., \& {Rabin}, D. 1992, \aap, 261, L21

\bibitem[{{Thomas}(1988)}]{Thomas1988}
{Thomas}, J.~H. 1988, \apj, 333, 407

\bibitem[{{Thomas} \& {Montesinos}(1991)}]{Thomas1991}
{Thomas}, J.~H. \& {Montesinos}, B. 1991, \apj, 375, 404

\bibitem[{{Uitenbroek} {et~al.}(2006){Uitenbroek}, {Balasubramaniam}, \&
  {Tritschler}}]{Uitenbroek2006}
{Uitenbroek}, H., {Balasubramaniam}, K.~S., \& {Tritschler}, A. 2006, \apj,
  645, 776

\bibitem[{Vernazza {et~al.}(1981)Vernazza, Avrett, \&
  Loeser}]{Vernazza+al:1981}
Vernazza, J.~E., Avrett, E.~H., \& Loeser, R. 1981, \apjs, 45, 635

\end{thebibliography}

\end{document}